\newcommand{\be}{\begin{eqnarray}}
\newcommand{\ee}{\end{eqnarray}}
\newcommand{\nn}{\nonumber}
\newcommand{\p}[1]{(\ref{#1})}
\newcommand{\lb}[1]{\label{#1}}
\newcommand\s{\scriptscriptstyle}

\newcommand\q{\quad}

\newcommand\Tr{\mbox{Tr}\,}

\def\a{\alpha}
\def\da{{\dot\alpha}}

\def\g{\gamma}

\def\d{\delta}

\def\eps{\epsilon}

\def\ve{\varepsilon}

\def\o{\omega}

 \def\th{\theta}

\def\L{\Lambda}

\def\pa{\partial}

\newcommand\A{{\s A}}

\newcommand\M{{\s M}}

\newcommand{\pp}{{++}}
\newcommand{\m}{{--}}

\newcommand{\Dp}{D^{\pp}}
\newcommand{\Dm}{D^{\m}}

\newcommand{\dpp}{\partial^{\pp}}
\newcommand{\dm}{\partial^{\m}}

\newcommand{\Vp}{V^\pp}
\newcommand{\Vm}{V^\m}

\def\sfrac#1#2{{\textstyle\frac{#1}{#2}}}

\documentclass[12pt]{article}

\usepackage{amsmath,amssymb}
\usepackage{graphicx}
\usepackage{cite}
\usepackage{bm}

\def\theequation{\arabic{section}.\arabic{equation}}
\begin{document}

\renewcommand{\thefootnote}{\fnsymbol{footnote}}

\vskip 15mm

\begin{center}

{\Large Renormalizable supersymmetric gauge theory in six dimensions}

\vskip 4ex

E.A. \textsc{Ivanov}$\,^{1}$,
A.V. \textsc{Smilga}\,$^{2}$,
B.M. \textsc{Zupnik}$\,^{1}$

\vskip 3ex

$^{1}\,$\textit{ Bogoliubov  Laboratory of Theoretical Physics, JINR, 141980 Dubna,
Russia}
\\
\texttt{eivanov, zupnik@theor.jinr.ru},
%\texttt{zupnik@theor.jinr.ru}
\\[3ex]
$^{2}\,$\textit{SUBATECH, Universit\'e de
Nantes,  4 rue Alfred Kastler, BP 20722, Nantes  44307, France
\footnote{On leave of absence from ITEP, Moscow, Russia.}}
\\
\texttt{smilga@subatech.in2p3.fr}
\end{center}

\vskip 5ex

\begin{abstract}
\noindent We construct and discuss a $6D$  supersymmetric
gauge theory involving
four derivatives in the action.
The theory involves  a dimensionless coupling
constant and is renormalizable. At the tree level, it
enjoys ${\cal N}{=}(1,0)$ superconformal symmetry,
but the latter is broken by quantum anomaly.
Our study should be considered as preparatory for seeking
an extended version of this theory which would hopefully preserve conformal symmetry at the
full quantum level and be ultraviolet-finite.
\end{abstract}

\renewcommand{\thefootnote}{\arabic{footnote}}
\setcounter{footnote}0
\setcounter{page}{1}

\section{Introduction}
Higher-dimensional quantum field theories bear interest from different
points of view and appear in numerous intertwining contexts, such as
Kaluza-Klein approach, string theory, higher spin theory, etc.

Recently, one of the present authors suggested \cite{duhi} that some
field theory in higher dimensions could play a role of
fundamental microscopic theory. This hypothetical underlying higher-dimensional
theory should, in particular, involve 3-brane classical solutions, which
might be associated with our Universe in the spirit of \cite{Rub}. In contrast to other
popular brane-Universe scenarios, like
Randall-Sundrum scenario \cite{BW}, the fundamental theory of the bulk in this case
is not assumed to include gravity, the latter is expected to be
generated as an effective theory living on the brane.
Clearly, there should exist a mechanism of getting rid of the cosmological term
which is known to be zero or very small.
For ensuring this, the fundamental theory
should be supersymmetric. Indeed, only supersymmetry
can provide for the exact cancellation of quantum
corrections to the energy density of the brane solution.

If we want the ``ultimate'' higher-dimensional theory to be renormalizable,
the canonical dimension of the lagrangian
should be greater than 4, i.e. it should involve higher derivatives.
Higher derivative theories are known to have a problem of
ghosts, which in many cases break unitarity and/or causality of the theory.
However, a model study performed in Refs. \cite{duhi}
indicated that in some cases, namely, when the theory enjoys {\it exact}
conformal invariance,  the ghosts are not
so malignant and the theory might enjoy a unitary S-matrix to any order of
perturbation theory.

We conclude that the conjectural fundamental QFT should preferably be a superconformal
theory. This restricts the number of  dimensions in the flat
space-time where the theory is formulated by $D \leq 6$. Indeed,
all standard superconformal algebras (involving the super-Poincar\'e algebra as a
subalgebra) are classified (for  instructive reviews see \cite{sconf}).
The highest possible dimension is six,
which allows for the minimal ${\cal N}{=}(1,0)$ conformal superalgebra
and the extended chiral ${\cal N}{=}(2,0)$ conformal superalgebra.

Thus a natural hypothesis is that the field theory in question lives
in six dimensions and enjoys the highest possible superconformal
(and super-Poincar\'e) symmetry with ${\cal N}{=}(2,0)$.
Unfortunately, no field theory with this symmetry group is known to
date. A possible candidate is the superconformal theory of tensor
(2,0) multiplet.\footnote{This multiplet is closely related to the
famous $M$-theory 5-brane \cite{M5,M52}.} However, the corresponding
lagrangian (with a standard, linear realization of ${\cal N}=(2, 0)$
superconformal symmetry\footnote{The nonlinear effective action of
(2,0) tensor multiplet as the world-volume multiplet of $M$-theory
5-brane, with nonlinearly realized ${\cal N}=(2,0)$ superconformal
symmetry, was constructed in \cite{M52}.}) is not constructed, and
only indirect results concerning scaling behavior of certain
operators have been obtained so far \cite{Sok}.

In this article, we derive the lagrangian for the $6D$ gauge theory with unextended
${\cal N}{=}1$ superconformal symmetry. This theory
is conformal at the classical level and renormalizable. However,
it is not finite: the $\beta$ function does not
vanish there and the conformal symmetry  is broken at the quantum level by
anomaly. In other words, the theory considered in this paper
cannot be regarded as a viable candidate for the ultimate theory. However,
its study represents a necessary preparatory step before tackling
the problem of constructing and studying a possible extension
of this theory, such that it would respect the superconformal
symmetry (at least the ${\cal N}{=}(1,0)$ one) at the full quantum level
and so could be considered as the appropriate candidate.

The adequate technique for constructing the relevant $6D$
superfields and their interactions is the technique of harmonic
superspace (HSS) \cite{HSS} which was extended to six dimensions in
\cite{HSW,Z1,Z2}. In the next Section we briefly describe the HSS
technique in six dimensions,  derive the lagrangian of
higher-derivative supersymmetric Yang-Mills theory and prove its
conformal invariance. In Sect. 3, we derive the lagrangian in the
component form. In Sect. 4, we calculate the $\beta$ function of
this theory at the 1-loop level. Its sign is positive so that the
theory has the Landau pole.\footnote{This fact might be understood
as follows. The ``usual'' non-gauge and Abelian gauge theories in
four dimensions have the Landau pole, while non-Abelian theories are
asymptotically free. On the other hand, a non-gauge  theory of the
real scalar field with cubic (not bounded from below) potential
$\propto \phi^3$ in six dimensions is known to be asymptotically
free \cite{phi3}. } The last Section is devoted, as usual, to
conclusions and speculations.

\section{$6D$ harmonic superspace.}
\setcounter{equation}0

 The basic facts about the spinor representations of
$SO(5,1)$ group and the ${\cal N}{=}1, 6D$ superspace can be found in Appendix.
What we actually need to know
is that the standard ${\cal N}{=}1$ superspace
(to be more precise, ${\cal N}{=}(1,0)$ superspace) involves the following coordinates
\be
z=(x^M, \th^a_i)\,,\q (M=0,...,5\,, \q a= 1,...,4\,, \q i = 1,2)\,,  \label{Centr}
\ee
where the Grassmann coordinates $\th^a_i$ obeys the reality condition
\be
\overline{\th^a_i}\equiv -C_{\dot{b}}^a(\th^b_i)^* =\th^{ai}\,.
\ee
Here the bar operation is the covariant conjugation defined in (\ref{covconj}).
The fact that $\bar \theta$ can be expressed via $\theta$ is a distinguishing feature of $6D$ superspace
compared to $4D$ superspace.

The basic spinor derivatives of the $6D, \,{\cal N}{=}1$ superspace are
\be
&&D^k_a=\pa^k_a - i\th^{bk}\pa_{ab}\,,\q\{D^k_a,D^l_b\}=-2i\ve^{kl}\pa_{ab}\,,
\ee
where
\be
&&\pa_{ab}=\sfrac12(\gamma^M)_{ab}\pa_M,\q \pa_M x^N=\delta^N_M,\q
\pa^k_a\th^b_i=\d^k_i\d^b_a\,, \q x^M = \sfrac12 (\gamma^M)_{ab}x^{ab}\,.
\ee

The off-shell superfield constraints of the $6D, \,{\cal N}{=}1$ gauge
theory have the following form \cite{HST,Ko}:
\be
\{\nabla^k_a,\nabla^l_b\}+\{\nabla^l_a,\nabla^k_b\}=0\,,\lb{constr}
\ee
where $\nabla^i_a=D^i_a+A^i_a(z)$ is the spinor covariant derivative;
$A^i_a$ is the spinor superfield
connection.

These constraints have been solved \cite{Z1,Z2} in the framework of the HSS approach.
To make the discussion self-contained and to establish the notation,
we describe it briefly here.

Let us first observe that the symmetry group of the superspace
$(x^M, \theta^a_i)$ involves
besides Poincar\'e and supersymmetry transformations also
$R$-symmetry $SU(2)$ transformations.
The conventional superspace is a coset of the
super-Poincar\'e transformations. It is natural to consider also the coset of the
$R$-symmetry   $SU(2)/U(1) = CP^1 \equiv S^2$.
It is parametrized by the  harmonics $u^{\pm i}$ ($u^-_i = (u^{+i})^*$,
$u^{+i}u^-_i = 1$).
The harmonic $6D, \;{\cal N}{=}1$ superspace is parametrized by the coordinates
$$
(z, u) = (x^M, \theta^a_i, u^{\pm i})\,.
$$
The harmonic superspace in the analytic basis involves the harmonics and
the so called analytic coordinates $Z_\A=(x_\A^M, \th^{\pm a})$
\be
x^M_\A=x^M+ \frac i2 \th^a_k\g^M_{ab}\th^b_l u^{+k}u^{-l},\q \th^{\pm a}=
u^\pm_k\th^{ak}.
\ee

It is convenient to define the following differential operators called
spinor and harmonic derivatives (in the analytic basis):
    \be
&& D^+_a=\pa_{-a}~, \q
D^-_a=-\pa_{+ a}-2i\th^{-b}\pa_{ab}\,, \q D^0 = u^{+i} \frac {\pa}{ \pa u^{+i}} -
u^{-i} \frac {\pa}{ \pa u^{-i}} + \th^{+a} \pa_{+ a} -  \th^{-a} \pa_{- a}
  \nn \\
&& \Dp=\dpp+i\th^{+a}\th^{+b}\pa_{ab}+\th^{+a}\pa_{-a}~,\q\Dm=\dm+i\th^{-a}
\th^{-b}\pa_{ab}+\th^{-a}\pa_{+a}~,
 \ee
where $\pa_{\pm a}\th^{\pm b} = \d^b_a $ and
$$ \dpp =  u^{+i} \frac {\pa }{ \pa u^{-i}} , \q \dm =  u^{-i} \frac {\pa }{ \pa u^{+i}}\ .
$$
The following commutation relations hold
 \be
&& \{D^+_a,D^-_b\}=2i\pa_{ab}, \q [\Dp, \Dm] = D^0 \nn \\
&& [\Dp,D^+_a]=[\Dm,D^-_a]=0\, ,\q[\Dp,D^-_a]=D^+_a,\q[\Dm,D^+_a]=D^-_a.
\ee
We shall use the notation
\be
(D^\pm)^4=-\sfrac{1}{24}\ve^{abcd}D^\pm_a D^\pm_b D^\pm_c D^\pm_d
\ee
and the following conventions for the full and analytic superspace
integration measures:
\be
\label{Zizeta}
d^{14}Z_\A=d^6x_\A\,(D^-)^4(D^+)^4,\q d\zeta^{-4}=d^6x_\A\,(D^-)^4.
\ee
 The following simple identity,
\be
\label{kBoxu}
\sfrac12(D^+)^4(\Dm)^2(D^+)^4=\Box(D^+)^4,\ \ \ \  \Box \equiv
\pa^\M\pa_\M=\sfrac12\ve^{abcd}\pa_{ab}\pa_{cd}\ ,
\ee
will be helpful for us.
\footnote{What we will actually need is the equivalence of the differential operators
$(1/2)(D^+)^4 (D^{--})^2$ and $\Box$ when acting on a {\it Grassmann-analytic}
(see below) superfield $\phi(\zeta, u)$.}

\subsection{Harmonic superfields and their interactions}

A general $6D$ superfield depends on 8 odd coordinates $\theta^a_i$ (or $\theta^{\pm a}$),
which makes their component expansion
rather complicated. There is, however, an important class of superfields,
{ Grassmann-analytic} superfields,
which depend only on
  \be
(\zeta, u) =(x^M_\A, \th^{+a}, u^{\pm i}) \, \label{AnalSS}
\ee
i.e. involves only half of the original Grassmann coordinates. The
set \p{AnalSS} forms the closed superspace on which
$6D, \,{\cal N}{=}(1,0)$ supersymmetry (and the full ${\cal N}{=}(1,0)$ superconformal symmetry)
can be realized and which is called ``harmonic analytic superspace''.
The structure of Grassmann-analytic (G-analytic) superfields is much simpler than
that of a general superfield.
The possibility to formulate the theory in terms of
G-analytic superfields represents a crucial advantage of the HSS formalism.
A certain disadvantage is that
the superfields depend now not only on the superspace coordinates,
but also on harmonics $u_i$.
The experience shows, however, that
the simplifications brought about by analyticity are more important
than the complications coming from
explicit harmonic dependence.
 A G-analytic superfield  $\phi(\zeta, u)$ satisfies the constraint
$D_a^+ \phi = 0$.\footnote{It is quite analogous to the habitual
chirality constraint $D_\alpha \phi = 0$ in four dimensions.}
In the analytic basis $D_a^+$ is reduced to the partial derivative
$\partial/\partial \theta^{- a}$ and this constraint simply means that
$\phi$ lives on the superspace \p{AnalSS}.

The superfields can be classified according to their harmonic charge $q$,
the eigenvalue of $D^0$. By the full analogy
with what is known for ${\cal N} = 2$ $4D$ theories \cite{HSS}, the $6D$
SYM theory is formulated   in terms of the  G-analytic
anti-Hermitian
superfield gauge potential which has charge $+2$ and is denoted $V^{++}$.
It defines the covariant harmonic derivative
\be
\nabla^\pp=\Dp+\Vp\,.
\ee

The superfield  gauge transformation uses the analytic anti-Hermitian matrix
parameter $\L$
\be
\d_\L\Vp= \Dp\L+[\Vp,\L]\,.\lb{gaugetran}
\ee

It is convenient to introduce also non-analytic gauge connection $V^{--}$
which can be obtained out of
$V^{++}$ as a solution of  the harmonic zero-curvature
equation
\be
\lb{A2}
\Dp\Vm-\Dm\Vp+[\Vp,\Vm]=0\,.\lb{hzc}
\ee
 %In Abelian case (no commutator), the solution has a very simple form if choosing
 %the gauge $D^{++} V^{++} = 0$, which is a superfield counterpart of the Landau gauge.
%In  this case, we have
%  \be
% \label{VmmAb}
%V^{--}_{\rm Ab} \ =\ \frac 12 (D^{--})^2 V^{++}\ .
%  \ee
The connection $V^{--}$ can be constructed
as a series over products of $V^{++}$ taken at different
harmonic ``points'',
\be
\label{Vmmrjad}
\Vm(z,u)=\sum\limits^{\infty}_{n=1} (-1)^n \int du_1\ldots du_n\,
\frac{\Vp(z,u_1 )\ldots \Vp(z,u_n )}{(u^+ u^+_1)(u^+_1 u^+_2)\ldots (u^+_n u^+)}\,,
\ee
where the factors $(u^+ u^+_1)^{-1}$ etc are the harmonic
distributions \cite{HSS} and the central basis coordinates $z$ are
defined in \p{Centr}. The connection $V^{--}$  transforms as
 \be
\d_\L\Vm= \Dm\L+[\Vm,\L]
\ee
under gauge transformations.
 It  can be used to build up spinor and vector superfield connections
    \be
A^-_a(V)=-D^+_a\Vm,\q A_{ab}(V)=-\sfrac{i}{2}D^+_aD^+_b\Vm.
    \ee
  In addition, one can define the covariant (1,0) spinor superfield strength,
\be
W^{+a}=-\sfrac{1}{6}\ve^{abcd}D^+_b D^+_c D^+_d\Vm\, .
\ee
 The superfield action of the {\it standard} $6D, \, {\cal N}{=}1$ gauge
theory was constructed in
\cite{Z1},
\be
S=\frac{1}{f^2}\sum\limits^{\infty}_{n=1} \frac{(-1)^{n+1}}{n} \int
d^6\!x\, d^8\theta\, du_1\ldots du_n \frac{\Tr  \left\{\Vp(z,u_1 )
\ldots \Vp(z,u_n ) \right\}}{(u^+_1 u^+_2)\ldots (u^+_n u^+_1 )}\,,\lb{action1}
\ee
where $f$ is the coupling  constant  of canonical dimension -1. The corresponding
component Lagrangian in the Wess-Zumino gauge [see eq.(\ref{standcomp}) below]
 gives the standard equations of motion of the 2-nd order for the gauge fields
 and of the 1-st order  for the fermions.

Let us derive the superfield equation of motion
for the action \p{action1}. To this end, one should first represent
the action as an integral over the analytic
superspace by acting with the operator $(D^+)^4$ on the integrand [cf.  (\ref{Zizeta})].
Taking
the variation of the result over $\delta V^{++}$ and comparing it with (\ref{Vmmrjad}),
we obtain
the equation
\be
\label{Fpp=0}
F^\pp=\sfrac14 D^+_aW^{+a}=(D^+)^4 \Vm=0~.\lb{eq1}
\ee
%In the superfield Landau gauge $\Dp\Vp=0$,
%the linear in  $\Vp$ term of this equation has the form
%     \be
%    F^\pp=\Box \Vp +O[(\Vp)^2]\,,
%    \ee
%as follows from (\ref{kBoxu}).

The superfield $F^{++}$ is Grassmann-analytic. It is transformed as
   \be
  \label{FppLam}
\delta_\Lambda F^{++} = [F^{++}, \Lambda]
  \ee
 under gauge
transformations.

%\vspace{.2cm}
It is very easy to write down now the superfield action with
{\it dimensionless} coupling constant. It has the following form
 \be
\label{dejstvie}
S = \frac{1}{2g^2}\int  d\zeta^{-4}du \, \Tr\left(F^\pp\right)^2\lb{hactan}\ .
\ee
Indeed, the superfields $V^{++}, V^{--}$ are dimensionless.
It follows that $F^{++}$ defined in (\ref{Fpp=0}) has canonical
dimension 2. Hence $g$ is dimensionless. The action (\ref{dejstvie})
is gauge invariant as follows immediately
from (\ref{FppLam}).

 The  action (\ref{dejstvie}) can be rewritten
as an integral over the full $6D$ harmonic superspace, in a few equivalent forms.
The corresponding Lagrangians are  the Chern-Simons type densities,
    \be
   S &=&  \frac{1}{2g^2}\int d^{14}Z_\A du\, \Tr \left(\Vm F^\pp\right)=
%   \frac{1}{48g^2}\int d^{14}Z_\A du\,\ve^{abcd} \Tr \left(D^+_a \Vm
%   D^+_bD^+_cD^+_d\Vm\right)\nn\\
   \frac{1}{8g^2}\int d^{14}Z_\A du\,\Tr \left(A^-_\a W^{+a}\right) \nn \\
&=& \frac{1}{12g^2}\int d^{14}Z_\A du\,\ve^{abcd}\Tr \left[ A_{ab}(V)A_{cd}(V)\right].
   \lb{hactgen}
   \ee

Note also that one can use in this model the alternative formalism with
an auxiliary tensor superfield $H^\pp$
\be
S(F^\pp,H^\pp)=-\frac{1}{g^2}\int  d\zeta^{-4}du\,\Tr\left[F^\pp(\Vp) H^\pp
 +\sfrac12\Tr(H^\pp)^2\right]\lb{FHact}
\ee
which is completely equivalent to the higher-order formalism.

To derive the  equations of motion one should use the
following tensor relation between arbitrary variations of harmonic
connections:
\be
\d\Vm &=&\sfrac12(\nabla^\m)^2\d\Vp
- \sfrac12 \nabla^{++}(\nabla^{--}\delta V^{--}), \nn\\
\q\nabla^\m\d\Vp &=& \Dm\d\Vp+[\Vm,\d\Vp]\,,
\label{Rel1}
\ee
or equivalently
\be
 \nabla^\m\d\Vp=\nabla^\pp\d\Vm,\q[\nabla^\pp,\nabla^\m]=D^0\,. \label{EquivForm}
\ee
The formula for $\delta V^{--}$ can be obtained by applying $\nabla^{--}$
to both sides of the first relation in \p{EquivForm} and using the second relation.
The variation of $S$ is
\be
\d S&=&\frac{1}{g^2}\int  d\zeta^{-4}du\,  \Tr\left(\d F^\pp\,F^\pp\right)
= \frac{1}{g^2}\int d^{14}Z_\A du\, \Tr \left(\d\Vm \,F^\pp\right)  \nn  \\
&=& \frac{1}{2g^2}\int d^{14}Z_\A du\, \Tr \left[\left(\nabla^\m\nabla^\m\d\Vp\right)\,
F^\pp\right]   \nonumber \\
&=& \frac{1}{4g^2}\int d\zeta^{-4} du\, \Tr \left[\d\Vp  (D^+)^4(\nabla^\m)^2(D^+)^4\Vm\right]   =0\,,
\ee
which leads to the equation
  \be
(D^+)^4(\nabla^\m)^2(D^+)^4\Vm=0\, .
\lb{highequ}
\ee
  Note that in the process of deriving this equation, the second term in the formula
for the variation $\delta V^{--}$ in \p{Rel1} was omitted since it does not contribute
by virtue of the important relation
\be
\nabla^{++}F^{++} = 0\,, \label{Vazhnoe}
\ee
which follows from the definition of $F^{++}$ in \p{eq1} and the harmonic
zero-curvature condition \p{hzc}.

\subsection{Superconformal invariance}
% EJ BOGU - LUCHSHE V APPENDIX !
 The action (\ref{dejstvie}) is scale invariant which suggests
its conformal invariance. In this subsection we prove it
within the superfield HSS formalism.

Transformations of the $6D, \,{\cal N}{=}(1,0)$ superconformal group $OSp(8^*|2)$
in the central basis have the form
\be
\d x^{ab}&=&c^{ab}+\o^a_cx^{cb}+\o^b_cx^{ac}+a\,x^{ab}
-\sfrac{i}{2}(\eps^{ak}\th^b_k-\eps^{bk}\th^a_k)\nn\\
&&+\,x^{ac}k_{cd}x^{db}+\sfrac14\th^{ak}\th^{bl}\th^c_l\th^d_kk_{cd}
-\sfrac{i}{2}\eta^k_c\th^a_kx^{bc}+
\sfrac{i}{2}\eta^k_c\th^b_kx^{ac} \nn\\
&&+\,\sfrac14\eta_{dl}\th^{bl}\th^{ak}\th^d_k-
\sfrac14\eta_{dl}\th^{al}\th^{bk}\th^d_k\,,\nn\\
\d\th^a_k&=&\eps^a_k+\o^a_b\th^{b}_k+\sfrac12 a\th^a_k-L^l_k\th^a_l
+x^{ac}k_{cd}\th^d_k+\sfrac{i}{2}\th^{al}\th^c_l\th^d_kk_{cd}\nn\\
&&-\,x^{ac}\eta_{ck}+i\eta_c^l\th^a_l\th^c_k-
\sfrac{i}{2}\eta_{ck}\th^{al}\th^c_l\,.\label{SconfCentr}
\ee
The meaning of the group parameters is clear from their
index structure and dimensions. In particular, $a$ is the dilatation
parameter, $k_{ab}$  are the parameters of special conformal transformations,
$\epsilon^{ak}$ and $\eta^k_a$ are the parameters
of $6D$ Poincar\'e and special conformal supersymmetries, $L^l_k$ are the parameters of $SU(2)$ rotations.
The closeness of the transformations \p{SconfCentr} can be directly checked.

The  conformal transformation of the harmonics
can be defined by the analogy with \cite{HSS}
\be
\delta u^+_k&=&\Lambda^\pp u^-_k~,\q\delta u^-_k=0\,,\nn \\
\Lambda^\pp&=&ik_{ab} \th^{+a}\th^{+b} +2i\eta^+_a\th^{+a}+
L^{kl}u^+_ku^+_l\,, \nn\\
\Dm\L^\pp&=&2ik_{ab}\th^{-a}\th^{+b}+2i\eta^-_a\th^{+a}+2i\eta^+_a\th^{-a}
+2L^{kl}u^-_ku^+_l\,, \label{Sconfharm}
\ee
where $\eta^{+}_a = \eta^i_au^+_i$. They have the same closure as \p{SconfCentr}.
The superconformal transformations of the harmonic
derivatives are given by
\be
\delta\Dp=-\Lambda^\pp D^0~,\q \delta\Dm=-(\Dm\Lambda^\pp)\Dm\,,
\ee
like in the $4D$ case \cite{HSS}.

Having the above transformations at hand, it is not difficult to
find how the ${\cal N}{=}(1,0)$ superconformal group is realized in
the analytic basis \be \delta\theta^{+a}&=&\epsilon^{+a}+\sfrac12
a\theta^{+a}+\omega^a_b \theta^{+b}+L^{+-}\th^{+a}+x^{ac}_\A
k_{cd}\th^{+d}-x^{ab}_\A\eta^+_b
+i\eta^-_b\th^{+b}\th^{+a}\,, \nn\\
\delta x^{ab}_\A&=&c^{ab}+\o^a_cx^{cb}_\A+\o^b_cx^{ac}_\A+a
x^{ab}_\A+ i(\epsilon^{-a}\theta^{+b}-\epsilon^{-b}\theta^{+a})
-k_{cd} x_\A^{ac} x_\A^{bd} \nn\\
&&-\,i\eta^-_cx^{ac}_\A\th^{+b}+i\eta^-_cx^{bc}_\A\th^{+a}\,,\nn \\
\delta\theta^{-a}&=&\eps^{-a}+\sfrac12 a\theta^{-a}+\omega^a_b
\theta^{-b}+L^{--}\th^{+a}-L^{-+}\th^{-a}+(x^{ac}_\A-i\th^{-a}\th^{+c})\th^{-d}k_{cd}\nn\\
&&-\,x^{ac}_\A\eta^-_c
+i\eta^-_c\th^{-c}\th^{+a}-i\eta^+_c\th^{-c}\th^{-a}-i\eta^-_c\th^{+c}
\th^{-a}\,.\label{SconfAnal}
\ee
Here $L^{--} = L^{kl}u^-_iu^-_k$\,, etc. Using these transformation rules,
it is easy to establish the transformation
of the analytic superspace integration measure
$d\zeta^{-4}du = d^6 x_\A d^4\th^+ du$:
\be
\d( d\zeta^{-4}du )&=&
(\,\pa_{ab}\d x^{ab}+\pa^\m\L^\pp-\pa_{+a}\d\th^{+a}\,)(d\zeta^{-4}du)
\equiv 4\L d\zeta^{-4}du\,, \nn \\
4\L&=& 4a-2x^{ab}k_{ab}
-4i\eta^-_a\th^{+a}-2L^{+-}\,,\label{AnalM}
\ee
where we used
\be
\pa_{ab}\d x^{ab}&=&6a-3x^{ab}k_{ab}-3i\eta^-_a\th^{+a}\,, \;
\pa_{+a}\d\th^{+a}= 2a + 4L^{+-} - x^{ab}k_{ab} +3i\eta^-_a\th^{+a}\,,\nn \\
\pa^\m\L^\pp&=&2i\eta^-_a\th^{+a}+2L^{+-}\,,\q \Dp\L=-\sfrac12\L^\pp\,. \label{Aux}
\ee
Under the superconformal transformations given above the gauge potentials
$V^{\pm\pm}$ transform as
\be
\delta V^{++} = 0\,, \quad \delta V^{--} = -(D^{--}\Lambda ^{++}) V^{--}\,, \label{Vconf}
\ee
which mimics the transformation rules of the harmonic derivatives.
The defining harmonic zero-curvature equation \p{hzc} is manifestly covariant
with taking into account the relation
\be
D^{++}\Lambda^{++} = 0\,.
\ee

Let us now verify the superconformal invariance of the action \p{hactan}.
The invariance of \p{hactan} under dilatations with the parameter $a$ is evident. The
invariance under the $SU(2)$ transformations (with parameters $L^{(ik)}$)
can be checked using the transformation properties in the analytic basis
\be
\delta V^{--} &=& -2L^{+-} V^{--}\,, \; \delta (D^{+})^4 = 4L^{+-}(D^{+})^4\,, \;
\delta F^{++} = 2L^{+-}F^{++}\,, \nn \\
\delta (d\zeta^{-4}du) &=& -2 L^{+-}\,(d\zeta^{-4}du)\,. \label{Ltransf}
\ee
It is straightforward to obtain
\be
\delta_L S \sim \int d\zeta^{-4}du\, 2 L^{+-}\Tr \left(F^{++}\right)^2.
\ee
Next we represent $2L^{+-} = D^{++}L^{--}$ and integrate by parts to rewrite $\delta_L S $ as
\be
\delta_LS \sim -2\int d\zeta^{-4}du\, L^{--} \Tr\left(\nabla^{++}F^{++} F^{++}\right).
\ee
This expression is vanishing as a consequence of the relation \p{Vazhnoe}.

Let us now prove the invariance under the special conformal supersymmetry
with the parameters $\eta^i_a$.
Since all other superconformal transformations are contained in
the closure of this supersymmetry
with itself and with the Poincar\'e supersymmetry and the action is manifestly
invariant under the latter, the invariance
under the conformal supersymmetry
actually amounts to the invariance under the full superconformal group.

In the analytic basis, the covariant derivative $D^{+}_a = \partial_{-a}$
transforms as
\be
\delta_\eta D^+_a = -\frac{\partial \delta \theta^{-b}}{\partial \theta^{-a}} D^{+}_b =
i(\eta^+\cdot \theta^- + \eta^-\cdot \theta^+) D^{+}_a + i\eta^-_a\left(\theta^+\cdot D^+ \right) -
i\eta^+_a\left(\theta^-\cdot D^+ \right).
\ee
Then it is straightforward to find
\be
\delta_\eta (D^+)^4 = i(3\eta^+\cdot \theta^- + 5\eta^-\cdot \theta^+) (D^+)^4 +
\frac{i}{2}\epsilon^{abcd}\eta^+_a D^+_bD^+_cD^+_d\,.
\ee
Taking into account that for the considered case
\be
D^{--}\Lambda^{++} = 2i(\eta^-\cdot \theta^+ + \eta^+\cdot \theta^-)
\ee
and using the transformation law \p{Vconf} of $V^{--}$,
it is also straightforward to compute that
\be
\delta_\eta F^{++} = i(\eta^+\cdot \theta^- + 3 \eta^-\cdot \theta^+) F^{++} +
\frac{i}{6}\epsilon^{abcd}\eta^+_a D^+_bD^+_cD^+_d V^{--}\,. \label{varF1}
\ee
Despite the presence of two terms which, being taken separately,
break analyticity, it is easy to check
that this variation is still implicitly analytic: acting on
it by $D^{+}_a$ yields zero. Actually,
it can be given the following manifestly analytic form
\be
\delta_\eta F^{++} = 3i (\eta^-\cdot \theta^+) F^{++}
+ i (D^+)^4\left[(\eta^+\cdot \theta^-)\,V^{--}\right]. \label{varF2}
\ee
The analytic superspace integration measure is transformed as
\be
\delta_\eta (d\zeta^{-4} du) = -4i (\eta^-\cdot \theta^+)\, (d\zeta^{-4}du)\,,
\ee
then the variation of the action \p{hactan}, up to the overall renormalization factor,
is as follows
\be
\delta_\eta S &\sim & \int d\zeta^{-4} du\, (D^+)^4\left[2i\left(\eta^+\cdot \theta^-
+ \eta^-\cdot \theta^+\right)\Tr \left(V^{--}F^{++}\right)\right] \nn \\
&=& 2i \int d^{14}Z du \left(\eta^+\cdot \theta^-
+ \eta^-\cdot \theta^+\right)\Tr\left(V^{--}F^{++}\right).
\ee
Then we represent
\be
\eta^+\cdot \theta^- + \eta^-\cdot \theta^+ = D^{++} (\eta^-\cdot\theta^-)\,,
\ee
integrate by parts with respect to $D^{++}$ and use
the relations \p{Vazhnoe} and \p{hzc} in the form
$$
\nabla^{++} V^{--} = D^{--}V^{++}\,.
$$
After these manipulations the variation acquires the form
\be
\delta_\eta S \sim - 2i \int d^{14}Z du\, \left(\eta^-\cdot \theta^-\right)
\Tr\left(D^{--}V^{++} F^{++}\right).
\ee
Now one should again take off $(D^+)^4$ from the measure
and apply it to the integrand. Clearly,
when all four derivatives hit the expression under the trace, the result is zero.
The only extra terms appear
when one of the spinor derivatives hits $\theta^-$
(and yields a Cronecker symbol) while three remaining
ones hit $D^{--}V^{++}$ under the trace.
It is clear that the result is also vanishing. Thus we obtain the desired
result
\be
\delta_\eta S = 0\,.
\ee
Below we shall independently check that the component action is conformally invariant, which,
together with the invariance under the Poincar\'e supersymmetry, also implies
full superconformal invariance.

\section{Component action}
\setcounter{equation}0
We now derive the component form of the action (\ref{dejstvie}).

We start from the following component expansion of $V^{++}$
written in the Wess-Zumino gauge
  \be
\label{VppWZ}
V^{++}_{(WZ)} = \theta^{+a}\theta^{+b} A_{ab} +
2\sqrt{2} (\theta^+)^3_a \psi^{-a} -
3 (\theta^+)^4 {\cal D}^{--}\,, \label{WZg} \,
\ee
 where
\be
(\theta^+)^3_d &=& \sfrac{1}{6}\epsilon_{abcd} \theta^{+a}\theta^{+b}\theta^{+c}\,, \quad
(\theta^+)^4 = - \sfrac {1}{24}  \epsilon_{abcd}
\theta^{+a}\theta^{+b}\theta^{+c}\theta^{+d}\,,\nn \\
\psi^{-a}&=&\psi^{ai}u^-_i\,, \quad {\cal D}^{--} = {\cal D}^{ik}u^-_iu^-_k\,.
\ee
The expansion (\ref{VppWZ}) involves only the physical fields: gauge fields $A_M$  (remind that
$A_{ab} = \sfrac12 A_M (\gamma_M)_{ab}$),  gluino fields $\psi^{ai}$
and a $SU(2)$ triplet of the scalar fields
${\cal D}^{ik} = {\cal D}^{ki}$.
The particular  numerical coefficients in (\ref{VppWZ}) were introduced for further convenience.

If reducing the theory to 4 dimensions, we arrive at the
${\cal N}{=}2$ vector multiplet, which can also be represented
as the combination of the  ${\cal N}{=}1$  vector multiplet
and the adjoint chiral multiplet. The $6D$ gauge field gives
the $4D$ gauge field and a complex scalar, a $6D$ gluino field
is split in two $4D$ gluinos and the field ${\cal D}^{ik}$
is decomposed as
 \be
\lb{Dmatr}
 {\cal D}^{ik} \ =\ \left( \begin{array}{cc} \bar F & -D \\ -D & -F  \end{array} \right),
 \ee
where real $D$ is the auxiliary field of the ${\cal N}{=}1$
vector multiplet and complex $F$ is the auxiliary field
of the  adjoint chiral multiplet. $D$ and $F$ are auxiliary (i.e. non-dynamical)
fields just in the theory based on the {\it standard action} (\ref{action1})
(and its 4-dimensional counterpart) because they enter it
without derivatives. We will see that in the action (\ref{dejstvie})
they become {\it dynamical}.

To find the component action, one must solve \p{A2} with $V^{++}_{(WZ)}$
for $V^{--}$, act on that by $(D^+)^4$ to
find $F^{++}$,  substitute
the latter into \p{dejstvie}, and integrate the result over Grassmann and
harmonic variables. The calculations are tedious (mainly because
$V^{--}$ needed at the intermediate steps is not G-analytic),
but feasible.

  For solving
\p{A2} we decompose $V^{--}$ with respect to $\theta^{-a}$ with
 coefficients representing G-analytic superfields:
    \be V^{--} = v^{--} + \theta^{-b}v^-_b +
    \theta^{-c}\theta^{-d}v_{cd} + (\theta^-)^3_d v^{+d} +
   (\theta^-)^4v^{++}
   \label{thetaminusdec}
   \ee
and rewrite \p{A2} as
a set of rather cumbersome harmonic equations for the
coefficients. This set is as follows
\be
&& D^{++}v^{--} +
\theta^{+b}v^-_b + \theta^{+a}\theta^{+b}[A_{ab}, v^{--}] +
2\sqrt{2} (\theta^+)^3_d[\psi^{-d}, v^{--}]\nn \\
&& \q -3 (\theta^+)^4[{\cal D}^{--}, v^{--}]  = 0\,,\nn \\
&& D^{++}v^-_b + 2\theta^{+c}( v_{cb} - A_{cb})
 + \theta^{+a}\theta^{+c}[A_{ac}, v^-_b]
 - \epsilon_{bacd}\theta^{+a}\theta^{+c}\psi^{-d} - 3(\theta^+)^3_b {\cal D}^{--} \nn \\
&& \q +\, 2\sqrt{2} (\theta^+)^3_d\{\psi^{-d}, v^-_b\}
-3 (\theta^+)^4[{\cal D}^{--}, v^-_b] = 0\,, \nn \\
&& D^{++}v_{cd} + \frac12\epsilon_{cdab}\theta^{+a} v^{+b}
-i \theta^{+a}\theta^{+b} \nabla_{cd}A_{ab}
 - 2 i\sqrt{2}(\theta^+)^3_b  \nabla_{cd}\psi^{-b} \nn \\
&& \q +\, 3i (\theta^+)^4  \nabla_{cd}{\cal D}^{--}   =0\, ,
\nn \\
&& D^{++}v^{+a} - \theta^{+a}v^{++} +
\theta^{+b}\theta^{+c}[A_{bc}, v^{+a}]
+ 2\sqrt{2} (\theta^+)^3_b\{\psi^{-b}, v^{+a}\} \nn \\
&& \q -\,3 (\theta^+)^4 [{\cal D}^{--},v^{+a}] = 0\,, \nn \\
&& D^{++}v^{++} + \theta^{+a}\theta^{+b} [A_{ab}, v^{++}] + 2\sqrt{2} (\theta^+)^3_a [\psi^{-a}, v^{++}]
-3  (\theta^+)^4
[{\cal D}^{--}, v^{++}] = 0\, \label{EQS}
\ee
with $\nabla_{ab} = \partial_{ab} - i [v_{ab}, \cdot]\,$.
One solves these
equations by decomposing the G-analytic coefficients with respect to
$\theta^+$ and solving the resulting harmonic equations for the
component fields. This procedure is rather boring, but
straightforward. Actually, for constructing the action we need only
the highest component $v^{++}$ in the expansion  \p{thetaminusdec}
\be
v^{++} = (D^+)^4 V^{--} = F^{++} = \lambda^{++} +
\theta^{+a}\lambda^+_a + \theta^{+a}\theta^{+b}\lambda_{ab} +
(\theta^+)^3_a\lambda^{-a} + (\theta^+)^4\lambda^{--}\,.
\label{Fppresh}
\ee
For the component fields in \p{Fppresh} we obtain
the following expressions
\be
\lambda^{++} &=& - {\cal D}^{++}\,, \quad \lambda^{+}_a =
i\sqrt{2}\,(\gamma^M)_{ab}\nabla_M \psi^{+b}\,, \nn \\
\lambda_{ab} &=& \frac{1}{2}(\gamma^M)_{ab}\left[ i\nabla_M {\cal D}^{+-}
+ \nabla^NF_{NM}\right] + \epsilon_{abcd}\{\psi^{-c}, \psi^{+d}\}\,, \nn\\
\lambda^{-a}&=& \sqrt{2} \nabla^2 \psi^{ai} u^-_i + i\sqrt{2} F_{MN}(\sigma^{MN})^a_b\psi^{bi}u^-_i -
\frac{4\sqrt{2}}{3}[\psi^{ai}, {\cal D}^l_i]u^-_l \nn \\
&& \q +\, \sqrt{2}
[\psi^{ai}, {\cal D}^{kl}] u^-_{(i}u^-_k u^+_{l)}\,, \nn \\
\lambda^{--} &=& -\nabla^2 {\cal D}^{--} -3 [{\cal D}^{--}, {\cal D}^{+-}]
-2 i \{\psi^-, \gamma^M\nabla_M\psi^-\}\,. \label{compF}
\ee
Here
\be
\nabla_M &=& \sfrac12 (\tilde\gamma)^{ab}\nabla_{ab} = \partial_M - i[A_M, \cdot]\,, \q
\nabla^2 = \nabla^M\nabla_M\,, \nn \\
F_{MN} &=& \partial_M A_N - \partial_N A_M - i[A_M, A_N]
\ee
and $\sigma^{MN}$ is defined in \p{sigmadef} (see Appendix).

As a warm-up, let us first reproduce   the known component expression
for the standard $6D$ SYM  action (\ref{action1}).
 From (\ref{action1}),
(\ref{Vmmrjad}) and (\ref{Zizeta}), the quadratic
in fields part of the action can be represented as
 \be
\label{Squadr}
S^{\rm quadr} \ =\ -\frac 1{2f^2} \int d\zeta^{-4} d^6x du  \,
{\rm Tr}\left \{ V^{++} F_{\rm lin}^{++} \right\}.
  \ee
Multiplying (\ref{VppWZ}) by (\ref{Fppresh}) with the linearized
components \p{compF}, performing the Grassmann
and harmonic integrations using the
identities
 \be
\lb{tozhd}
&& \theta \gamma_M \theta  \theta \gamma_N \theta = - 8 \theta^4 \eta_{MN}\, , \q \int d\zeta^{-4}(\theta^+)^4 = \int d^6 x_A\,, \nn \\
&& \int du \, u^+_i u^-_j = \frac 12 \epsilon_{ij}, \ \ \ \ \ \ \ \ \
\int du \, u^+_i u^+_k u^-_j u^-_l = \frac 16 \left( \epsilon_{ij} \epsilon_{kl}
+  \epsilon_{il} \epsilon_{kj} \right),
 \ee
and restoring the nonlinear terms by gauge invariance, we obtain
  \be
 \lb{standcomp}
 S = \frac 1{f^2} \int d^6x  \, \mbox{Tr}
\left \{- \frac 12 F_{MN}^2 - \frac 12 {\cal D}^{ik}  {\cal D}_{ik} +
i \psi^k \gamma^M \nabla_M \psi_k
\right\}.
   \ee

The component form of the higher derivative action (\ref{dejstvie})
is also derived rather straightforwardly. After integrating over
$\theta^{+a}$, the action (\ref{dejstvie}) is expressed in terms
of the components of $F^{++}$, eq. (\ref{Fppresh}), as follows
\be
S = \frac{1}{2g^2} \int d^6x du \left(2 \lambda^{++}\lambda^{--}
 -2 \lambda^+_a\lambda^{-a} -
\epsilon^{abcd}\lambda_{ab}\lambda_{cd}\right). \label{Act1}
\ee
After substituting the expressions \p{compF} and performing
the integration over harmonics  in \p{Act1} we obtain the sought
component action:
  \be
 S &=& -\frac{1}{g^2}\int d^6x  \, \mbox{Tr}
\left\{ \left( \nabla^M F_{ML}\right)^2  +  i\psi^j\gamma^M \nabla_M (\nabla)^2\psi_j
 + \frac 12 \left(\nabla_M{\cal D}_{jk}\right)^2
\right. \nn \\
&& \q \left.
+\,  {\cal D}_{lk}{\cal D}^{kj}{\cal D}^{\;\;\;l}_{j}
  -2i {\cal D}_{jk} \left( \psi^j\gamma^M\nabla_M\psi^k
- \nabla_M \psi^j \gamma^M \psi^k \right)
+ (\psi^j\gamma_M \psi_j)^2  \right. \nn \\
&& \q \left. +\, \frac 12  \nabla_M\psi^j
\gamma^M\sigma^{NS}[F_{NS}, \psi_j]
- 2\nabla^M F_{MN}\, \psi^j\gamma^N\psi_j
 \right\}
\label{CompAct}
 \ee

Let us discuss this result.  Note first of all that the quadratic terms in the lagrangian
are obtained from \p{standcomp} by adding the extra box operator (it enters with negative sign,
this makes the kinetic time-derivative terms positive definite in Minkowski space). It is immediately seen
for the terms $\propto {\cal D}^2$ and for the fermions. This is true also for the gauge part due to the
identity
 \be
\lb{FboxF}
 {\rm Tr} \left\{ ( \nabla^M F_{MN} )^2 \right \} \
=\ -\frac 12 {\rm Tr} \left\{ F^{MN} \nabla^2 F_{MN} \right\}
-   2i {\rm Tr}\, \left\{ F_{M}^{\;\;\;N} F_{NS} F^{SM} \right\}.
 \ee
 The appearance of the structure (\ref{FboxF}) was anticipated in \cite{duhi}
by lifting a $4D$ higher-derivative
supersymmetric gauge lagrangian to six dimensions. Actually, only this higher-derivative
gauge fields kinetic term is conformally invariant.
\footnote{The four-derivative conformally invariant kinetic term
of gauge fields was also considered in \cite{Kaz} where it appeared
as an effective Lagrangian in a $6D$ non-conformal theory with standard
two derivatives in the action. Its interpretation in our case is entirely
different: it is present at the level of the microscopic Lagrangian and
defines the free propagators of gauge fields.}

As promised, the fields ${\cal D}^{ik}$ become dynamical. They carry canonical
dimension 2 and their kinetic
term involves  two derivatives.
There is a cubic term $\propto {\cal D}^3$. This sector of the theory
reminds the renormalizable theory
$(\phi^3)_6$. Gauge and fermion fields have the habitual canonical dimensions
$[A_M] = 1,\ [\psi] = 3/2$. Their kinetic terms involve, correspondingly, 4 and 3 derivatives.
 The lagrangian involves also other interaction terms, all of them having the canonical
dimension 6.

It is instructive to evaluate the number of on--shell degrees of
freedom in this lagrangian. Consider first the gauge field. With the
standard lagrangian $\propto {\rm Tr} \{F_{MN}^2\}$, a
six--dimensional gauge field $A_M$ has  4 on--shell d.o.f. for each
color index. The simplest way to see this is to note that $A_0$
is not dynamical and we have to impose the Gauss law
constraint on the remaining 5 spatial variables.
For the higher-derivative theory, however, the
presence of two extra derivatives doubles the number of d.o.f. and
the correct counting is $2\times 5 = 10$ before imposing the Gauss law constraint
and $10 -1 = 9$ after that. In addition, there are 3
d.o.f. of the fields $D_{ij}$ and we have all together 12 bosonic
d.o.f. for each color index. The standard $6D$ Weyl fermion (with
the lagrangian involving only one derivative) has 4 on--shell degrees of
freedom. In our case, we have $4\times 3 = 12$ fermionic d.o.f.
 due to the presence of three derivatives in the kinetic
term. For sure, the numbers of bosonic and fermionic degrees of
freedom on mass shell coincide.

Our theory is conformally invariant at the classical level. This was
proven in the previous Section using superfield formalism. But
conformal invariance of the component lagrangian (\ref{CompAct}) can
be shown directly. As distinct from the consideration in the
previous Section, we will use here the formalism where the
coordinates do not change after the transformation and only the
fields do (the ``active'' form of the transformations). A special
conformal variation of a primary  operator $\Phi$ is
   \be
 \lb{confvar}
\delta_C^* \Phi = \left[(2x^M x^N - \eta^{MN} x^2 ) \pa_N \Phi
+ 2x_N (\eta^{MN} d + S^{MN}) \Phi \right] \epsilon_M
\, ,
 \ee
 where $d$ is the canonical dimension of $\Phi$ and $S^{MN}$
is the spin operator.\footnote{In fact, the same check of the conformal invariance can be
performed using the ``passive'' variations $\delta_C \Phi = 2x_N\epsilon_M(\eta^{MN}d + S^{MN})\Phi$,
$\delta_C\nabla_M = 2\left[ (x\cdot \epsilon)\nabla_M - x_M(\epsilon\cdot\nabla) +\epsilon_M (x\cdot \nabla)\right]$,
$\delta_C(d^6x) = -12 (x\cdot \epsilon) d^6x$, where ``$\cdot$'' denotes the scalar product.}
The transformation law
(\ref{confvar}) has the same form in space of any dimension. It is easy
to see that for a primary operator $O$
of canonical  dimension $D$, $  \int d^Dx \, \delta_C  O = 0\,$.
Thus, conformal invariance of the action would be proven
if the lagrangian density is transformed as in (\ref{confvar})
with respect to conformal transformations (with $d=6$).
This is a nontrivial  requirement and not any operator
of canonical dimension $D$ satisfies it (for a good review
see e.g. the lecture of R. Jackiw \cite{Jackiw}).
 An example of a $4D$ action
 which is scale invariant, but not conformally invariant is
$$ O = \frac {[(\partial_\mu \phi)^2]^2}{\phi^4}\,. $$
 The point is that the derivatives of primary
operators do {\it not} enjoy the same transformation
laws (\ref{confvar}) as the primary operators. For example, if
$\Phi$ is a scalar of dimension $d$, the variation of $\pa_M \Phi$
under a special conformal transformation is
     \be
     \lb{der=nekonf}
     \delta^*_C (\pa_M \Phi) \ =\ \delta^*_C (\pa_M \Phi) + 2 d \Phi \epsilon_M\,.
     \ee
One can check, however, that most of the terms in (\ref{CompAct})
are transformed, up to a total derivative,
 according to the law (\ref{confvar}). On the other hand, this is not true
 for the kinetic fermion terms and two last terms.
    Denoting the corresponding
terms in the braces  (with the attached signs and coefficients)
in the consecutive order as $(I),
(II)$ and $(III)$, we find that their variations contain, besides
the standard  pieces (\ref{confvar})  and  total derivative terms,
also some extra pieces
\be
\hat{\delta}(I) &=&
- 4 \mbox{Tr}\left \{ F^{MN} \psi^i \g_N \psi_i \right \} \epsilon_M\,, \nn \\
 \hat{\delta}\,(II) &=&
- 4 \mbox{Tr}\left \{ F^{MN} \psi^i \g_N \psi_i \right \} \epsilon_M\,, \nn \\
 \hat{\delta}\,(III) &=& 8 \mbox{Tr}\left \{ F^{MN} \psi^i \g_N \psi_i  \right \} \epsilon_M\,.
\label{ExtrA}
\ee
 In deriving (\ref{ExtrA}), we used the identity
\be
\gamma_M\sigma_{NS} + \sigma_{NS} \gamma_M = 2(\eta_{MS}\gamma_N - \eta_{MN}\gamma_S)\,.
\label{idgam}
\ee

We see that
 \be
\hat{\delta}\left[(I) + (II) + (III) \right] = 0\,,
\ee which proves
the conformal invariance of \p{CompAct}.\footnote{It is interesting to note
that there exists another independent conformal combination of
these terms, $ \hat{\delta}\left[(III) + 2(II)\right] = 0$.}
As the action is supersymmetric by construction, this
proves in a simple way  the full superconformal invariance
of the action (\ref{dejstvie}), \p{CompAct}.

\section{Charge renormalization}
\setcounter{equation}0
We proceed now to calculating (at the one--loop level)
the $\beta$ function of our theory. We will see
that it does not vanish, which means that conformal invariance of the classical action is broken
by quantum effects.

\begin{figure}[h]
   \begin{center}
 \includegraphics[width=4.0in]{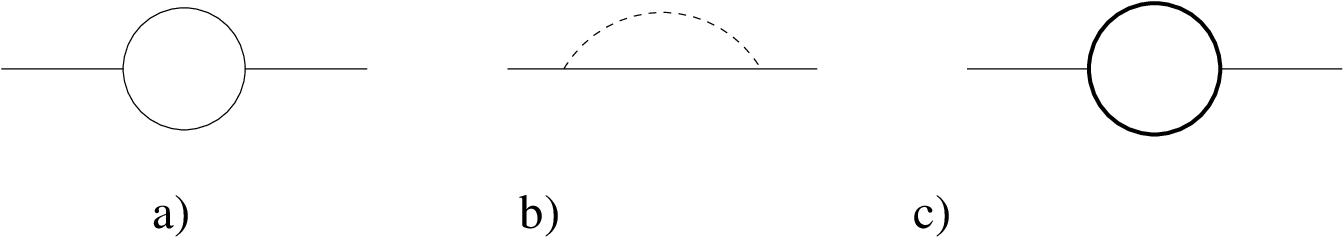}
        \vspace{-2mm}
    \end{center}
\caption{\small Graphs contributing to the renormalization of the kinetic term.
Thin solid lines stand for the particle
${\cal D}$, thick solid lines for fermions, and dashed lines for gauge bosons.}
\label{figD2}
\end{figure}

\begin{figure}[h]
   \begin{center}
 \includegraphics[width=4.0in]{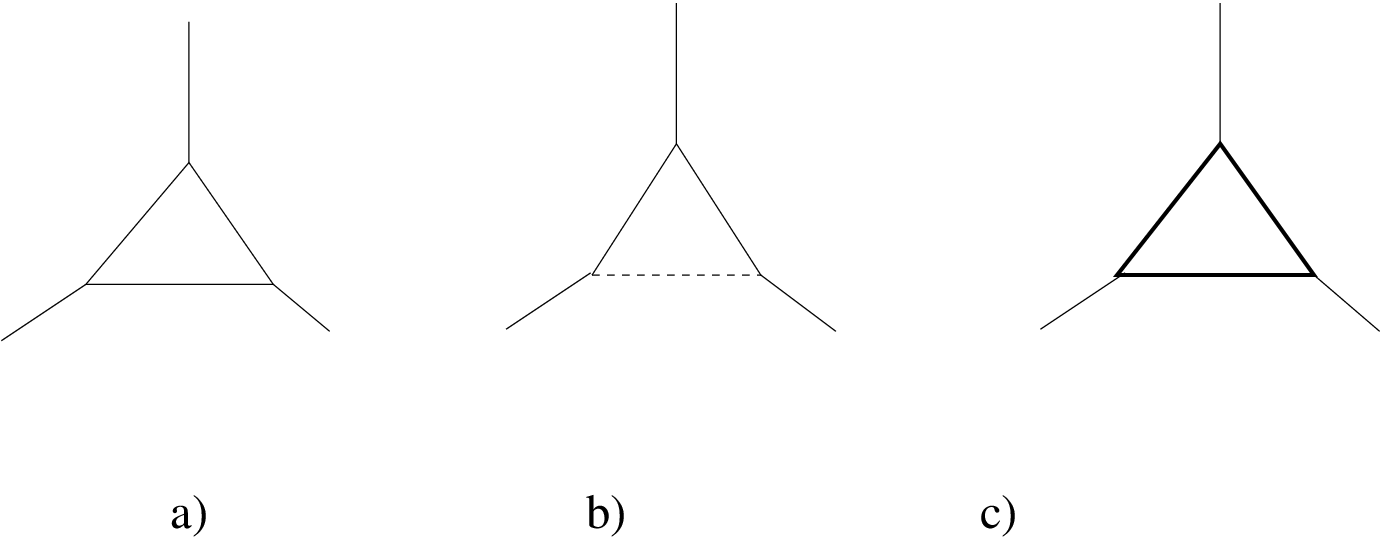}
        \vspace{-2mm}
    \end{center}
\caption{\small The same for the ${\cal D}^3$ vertex.}
\label{figD3}
\end{figure}

The simplest way to do this calculation is to evaluate 1--loop corrections
to the structures $\sim (\pa_M {\cal D})^2$
and $\sim {\cal D}^3$. The relevant Feynman graphs
are depicted in Figs. \ref{figD2}, \ref{figD3}.

For perturbative calculations, we absorb the factor $1/g$
in the definition of the fields. The relevant propagators are
  \be
\lb{prop}
 \langle A_M^A A_N^B \rangle  &=& - \frac {i \eta_{MN} \delta^{AB}}{p^4}\,, \nn \\
\langle \psi^{jA} \psi^{kB} \rangle &=&
- \frac {i \epsilon^{jk} \delta^{AB} p_N \tilde \g^N }{p^4}\,,    \nn \\
 \langle {\cal D}_{ik}^A {\cal D}_{jl}^B \rangle &=&
- \frac {i\delta^{AB}}{p^2} \left( \epsilon_{ij} \epsilon_{kl}
+  \epsilon_{il} \epsilon_{kj} \right),
 \ee
where $A,B$ are color indices, $A_M = A^A_M t^A\,$, $\Tr (t^At^B) = \delta^{AB}/2\,$, etc.
The vertices can be read out directly from the lagrangian.

 Consider first the graphs in Fig. \ref{figD2}.
They involve logarithmic and quadratic divergences. The latter do not appear if one uses dimensional regularization or any other regularization scheme that respects gauge invariance and supersymmetry. Indeed, the quadratically divergent pieces in Fig. \ref{figD2} are associated with the contribution $\sim {\rm Tr} \{D^2\}$ in the effective action. But this term is absent in the tree action \p{CompAct} and cannot appear in the effective one.

The logarithmic divergences in the 2-point graphs are
  \be
\Delta {\cal L}^{\rm eff}_{(2)} = g^2 c_V \left( - \frac 32 - \frac 76 +
4 \right)  {\rm Tr}\, \{(\pa_M {\cal D}_{jk})^2 \}\, L
= \frac {4g^2c_V}3  {\rm Tr}\, \{(\pa_M {\cal D}_{jk})^2 \}\, L\,,
  \ee
where\footnote{We use here the simple momentum cutoff as an ultraviolet regularization. This procedure does not respect the symmetries of the theory and would not assure, e.g., the cancellation of the quadratic divergences, but, as long as the logarithmic one-loop divergences are concerned, it gives the same result as more sophisticated methods, and its use is legitimate.
 The situation is exactly the same as in the ordinary QED where the momentum cutoff  gives a quadratically divergent photon mass, but assures a correct result for the one-loop logarithmic divergences in the effective charge.}
\be
 L = \int_\mu^\Lambda  \frac {d^6p_E}{(2\pi)^6 p_E^6}
= \frac 1{64\pi^3} \ln \frac \Lambda \mu
 \ee
and three terms in the parentheses correspond to the contributions
of the graphs in Fig. \ref{figD2}a,b,c.

The 3-point graphs in Fig. \ref{figD3} involve only logarithmic divergence. We obtain
  \be
\Delta {\cal L}^{\rm eff}_{(3)} = g^3 c_V \left( - \frac 92 + \frac 32 + \frac {32}3  \right)
{\rm Tr}\, \{ {\cal D}_{lk} {\cal D}^{kj} {\cal D}_j^l \}\, L
=  \frac {23 g^3 c_V}3  {\rm Tr}\, \{  {\cal D}_{lk} {\cal D}^{kj} {\cal D}_j^l \}\, L\,.
  \ee
 The full 1-loop effective lagrangian in the ${\cal D}$ sector is
   \be
{\cal L}^{\rm eff}_{{\cal D}} \ =\ - \frac 12
{\rm Tr}\, \{(\pa_M {\cal D}_{jk})^2 \} \left( 1 - \frac {8g^2c_V}3 L \right)
-g  {\rm Tr}\, \{  {\cal D}_{lk} {\cal D}^{kj} {\cal D}_j^l \}
\left( 1 - \frac {23 g^2c_V}3 L \right).
 \ee

Absorbing the renormalization factor of the kinetic term
in the field redefinition, we finally obtain\footnote{This result agrees with a recent calculation of L. Casarin and A. Tseytlin performed by a different method \cite{Casarin}.}
 \be
 \lb{rencharge}
g(\mu) \ =\ g_0 \left( 1 - \frac {11 g_0^2 c_V}3 L \right) = g_0
\left( 1 - \frac {11 g_0^2 c_V} {192 \pi^3} \ln
\frac \Lambda \mu \right)
 \ee
 for the effective charge renormalization.

The sign corresponds to the Landau zero situation, as in the ordinary 4-dimensional QED.
It is amusing to observe that,
if taking into account only the graphs in Fig. \ref{figD2}a
and Fig. \ref{figD3}a, the coefficient would be zero.
In other words, the purely bosonic $6D$ theory
 \be
{\cal L} \ =\  - \frac 12   {\rm Tr}\, \{(\pa_M {\cal D}_{jk})^2 \} -
g  {\rm Tr}\, \{  {\cal D}_{lk} {\cal D}^{kj} {\cal D}_j^l \}
  \ee
does not involve logarithmic divergences at the one--loop level.

\section{Conclusions}
In this paper we presented the first example of renormalizable
higher-dimensional supersymmetric gauge theory. It is $6D, {\cal
N}{=}(1,0)$ gauge theory with four derivatives in the action and
dimensionless coupling constant. Though it is superconformally
invariant at the classical level, the superconformal symmetry turns
out to be broken in the quantum case by conformal anomaly. As the
result of this breaking, in accord with the arguments of
\cite{duhi}, the quantum theory suffers from ghosts which cannot be
 entirely harmless.  This raises the problem
of searching for some extended theory, such that the
(super)conformal symmetry is retained in it at the full quantum
level. So this hypothetical theory would reveal the nice property of
ultraviolet finiteness and could probably be considered as a
candidate for the fundamental field theory. {\it Exact} conformal
invariance may render the ghosts harmless \cite{duhi}.

What would the extended theory look like?
We see only two options here:
 \begin{itemize}
\item
 It may enjoy the maximal superconformal symmetry ${\cal N}{=}(2,0)$
 in six dimensions. However, in this case it should
depend on tensor rather than vector multiplets \cite{M5,Sok}.
Unfortunately, to describe the tensor multiplet in the framework of
HSS  is not a trivial task and it is  not  solved yet. As a result,
no microscopic Lagrangian for interacting (2,0) tensor multiplet is
known today...
 \item
 Another possibility to try is to add to
the higher-derivative $6D$, ${\cal N}{=}(1,0)$ supersymmetric
gauge theory action \p{dejstvie} an off-shell action
of $6D, {\cal N}{=}(1,0)$ hypermultiplet in some representation
of the gauge group, with the same number of derivatives on fields.
A similar extension of the non-conformal standard
action of ${\cal N}{=}(1,0)$ gauge multiplet (having two derivatives) with
the hypermultiplet in the adjoint representation
is known to give rise to non-conformal ${\cal N}{=}(1,1)$ gauge theory \cite{HST}.
In the HSS approach, the hypermultiplet is described by an analytic superfield
$q^{+}(\zeta, u)$ which is unconstrained and so contains off shell
infinite towers of auxiliary fields with growing isospins (they come from
the harmonic expansions). The higher-derivative action of $q^+$ could hopefully
be constructed in such a way that it respects ${\cal N}{=}(1,0)$ superconformal
symmetry (which is surely broken in the standard minimal $6D$ hypermultiplet
action). However, the total higher-derivative ${\cal N}{=}(1,0)$ superconformal
action of the gauge multiplet and hypermultiplet cannot be expected to possess
neither ${\cal N}{=}(1,1)$ super Poincar\'e nor any extended superconformal
supersymmetry. Indeed, ${\cal N}{=}(1,0)$ gauge multiplet and hypermultiplet
can be combined only into a ${\cal N}{=}(1,1)$ gauge multiplet, while no superconformal
extension is known for non-chiral $6D, {\cal N}{=}(1,1)$ Poincar\'e supersymmetry \cite{sconf}.
Nevertheless, such a coupled system could hopefully preserve its
${\cal N}{=}(1,0)$ superconformal symmetry at the full quantum level,
thus being ultraviolet finite. This higher derivative $6D$ theory
should reveal some rather unusual features, since the infinite towers
of the ``former'' auxiliary fields collected in the harmonic expansion of $q^+$
should acquire kinetic terms and become propagating, like the ``former''
auxiliary field ${\cal D}^{ik}$ of the vector multiplet.
\end{itemize}

A less ambitious but important task is to repeat  the quantum calculations
of Sect. 4 with making use of the manifestly supersymmetric techniques of the harmonic
supergraphs \cite{HSS}. This would allow one to obtain the charge renormalization
for the whole superfield action \p{dejstvie}, i.e. at once for all fields of
the gauge multiplet. The supergraph techniques would be especially useful for
exploring the quantum properties of the hypothetical higher-derivative
gauge fields - hypermultiplet system.

\section*{Acknowledgements}

We are indebted to L. Casarin and A. Tseytlin, who detected an arithmetic error in the calculation of the coefficient of the charge renormalization in the earlier versions, and to I. Buchbinder, D. Sorokin,  M. Strassler and A. Vainshtein for useful discussions and comments.
E.I. and B.Z. acknowledge a support from RFBR grant, project  No
03-02-17440, NATO grant PST.GLG.980302, the grant INTAS-00-00254,
the DFG grant No.436 RUS 113/669-02, and a grant of the
Heisenberg-Landau program. E.I. thanks the SUBATECH, Nantes, for the
kind hospitality at the initial stages of this study.

\renewcommand\theequation{A.\arabic{equation}} \setcounter{equation}0
\section*{ Appendix: 6D, {\cal N}{=}1 superspace}

The group $Spin(5,1)$ has two different irreducible spinor representations of complex dimension 4.
An essential difference with the familiar  $Spin(3,1)$ case (that also involves two different spinor representations)
is that a complex conjugated $Spin(5,1)$ spinor belongs to {\it the same} representation of the group as the original one.
To see it explicitly, one can express
a (1,0) Weyl spinor $\Psi^a,~(a=1, 2, 3, 4)$ in terms of two $SL(2,C)$ spinors
   \be
  \Psi^a=\left(\begin{array}{l}\psi^\a\\
  \bar\kappa^\da\end{array}\right).
   \ee
The complex conjugation gives us
   \be
  (\Psi^a)^*\equiv \Psi^{\dot{a}}=\left(\begin{array}{l}\bar\psi^\da\\
  \kappa^\a\end{array}\right).
   \ee
This complex-conjugated representation of $Spin(5,1)$ is equivalent
to some covariant (1,0) spinor
   \be
 \Psi^{\dot{a}}=C^{\dot{a}}_a\bar\Psi^a,~\bar\Psi^a=-C^a_{\dot{a}} \Psi^{\dot{a}}\,,
  \ee
where the charge-conjugation matrix is introduced and
  \be
C^a_{\dot{a}}C^{\dot{a}}_b=-\delta^a_b\,.
  \ee

We shall use  the covariant conjugation
\be
\lb{covconj}
\Psi^a~\Rightarrow~\bar\Psi^a=\left(\begin{array}{l}-\kappa^\a\\
\bar\psi^\da\end{array}\right)
\ee
which has the unusual property $\overline{\bar\Psi^a}=-\Psi^a\,$.

We choose the  antisymmetric representation of the 6D Weyl matrices
\be
(\gamma^M)_{ab}=-(\gamma^M)_{ba}\,\q\tilde\gamma_M^{ab}=\sfrac12\ve^{abcd}
(\gamma_M)_{cd}\,,
\ee
where $M = 0,1,\ldots,5$ and $\ve^{abcd}$ is the totally antisymmetric symbol.
All these matrices are real with respect to the covariant conjugation
\be
\overline{(\gamma^M)_{ab}}=C^{\dot{c}}_a C^{\dot{d}}_b(\gamma^M_{cd})^*=
(\gamma^M)_{ab}\,.
\ee

The basic relations for these Weyl matrices are
\be
&&(\gamma_M)_{ac}(\tilde\gamma_N)^{cb}+(\gamma_N)_{ac}(\tilde\gamma_M)^{cb}
=-2\d^b_a
\eta_{MN}\,,\\
&&\ve_{abcd}=\sfrac12(\gamma^M)_{ab}(\gamma_M)_{cd}\,,
\ee
where $\eta_{MN}$ is the metric of the 6D Minkowski space
($\eta_{00}=-\eta_{11}=\ldots=-\eta_{55}=1)$ and $\gamma_M=\eta_{MN}\gamma^N\,$.

The generators of the (1,0) spinor representation are $S^{MN} =
-\frac12 \sigma^{MN}\,$, where \be
(\sigma^{MN})^b_a=\frac12(\tilde{\gamma}^M\gamma^N-
\tilde{\gamma}^N\gamma^M)^b_a\,,\
\overline{\sigma^{MN}}=\sigma^{MN}\,.\lb{sigmadef} \ee

\end{document}